\documentclass[aps,prl,twocolumn,groupedaddress,amsmath,amssymb]{revtex4}

\usepackage{graphics}

\newcommand{\dst}{\displaystyle}
\newcommand{\tst}{\textstyle}
\newcommand{\rhm}{\rho_{m}}
\newcommand{\prm}{p_{m}}
\newcommand{\dph}{\dot{\phi}}
\newcommand{\ddrhm}{\frac{d\phantom{\rho}}{d\rhm}}
\newcommand{\pat}{\partial}
\newcommand{\rhmz}{\rho_{m0}}
\newcommand{\prmz}{p_{m0}}

\begin{document}

\title{GAUGE APPROACH TO GRAVITATION AND REGULAR BIG BANG THEORY
}

\author{A. V. Minkevich}

\email[]{MinkAV@bsu.by; awm@matman.uwm.edu.pl}


 \affiliation{Department of Theoretical Physics, Belarussian State University,
av. F. Skoriny 4, 220050, Minsk, Belarus}
 \affiliation{ Department of Physics and Computer Methods, Warmia
and Mazury University in Olsztyn, Poland}

\date{\today}

\begin{abstract}
Field theoretical scheme of regular Big Bang in 4-dimensional physical space-time, built in the
framework of gauge approach to gravitation, is discussed. Regular bouncing character of homogeneous
isotropic cosmological models is ensured by gravitational repulsion effect at extreme conditions
without quantum gravitational corrections. The most general properties of regular inflationary
cosmological models are examined. Developing theory is valid, if energy density of gravitating
matter is positive and energy dominance condition is fulfilled.
\end{abstract}

\maketitle

\section{I. Introduction}

The Hot Bing Bang scenario built on a base of Friedmannian homogeneous isotropic cosmological models
(HICM) of general relativity theory (GR) is the foundation of  relativistic cosmology. Well known physical
laws allow to describe physical processes in evolving Universe beginning from the time $t_1\sim 10^{-4} s$
from the Bing Bang corresponding to the start of cosmological expansion (in the case of Friedmannian HICM
the moment $t=0$ corresponds to cosmological singularity), when the energy density of gravitating matter
was comparable with nuclear density and the temperature $T\sim 10^{11} K$. Observable confirmation of
theoretical predictions of primordial nucleosynthesis of light elements, explanation of large scale
structure formation of the Universe are important achievements of the Hot Bing Bang theory. To describe
gravitating matter and physical processes in evolving Universe at $t<t_1$ one has to use modern theory of
fundamental physical interactions, unified models of particle physics, that is the object of cosmology of
early Universe. The creation of inflationary paradigm, which permits to solve a number of problems of
standard cosmological scenario, in particular, to explain the homogeneity and isotropy of the Universe at
initial stages of cosmological expansion is important achievement of early Universe cosmology \cite{mc27}.
The further progress in cosmology of early Universe is connected with further perfection of particle
physics theory and also theory of gravitational interaction.

As it is known, GR has problems by description of gravitating matter at extreme conditions with extremely
high energy densities, pressures, temperatures (problem of gravitational singularities). The Friedmannian
HICM possess singular state in the past, in which some important physical characteristics of gravitating
matter (energy density, temperature etc) diverge and the scale factor $R(t)$ of Robertson-Walker metrics
vanishes. As a result the existence of Friedmannian HICM is limited in the time in the past. Corresponding
problem –- the problem of cosmological singularity (PCS) -- is one of the most principal cosmological
problems, which does not have general solution still. From physical point of view, the PCS is connected
with the fact that gravitational interaction in the case of usual gravitating systems with positive values
of energy density $\rho$ and pressure $p$ in the frame of GR as well as Newton's gravitation theory has
the character of attraction, but not repulsion. Although in the case of gravitating systems with negative
pressure (for example, massive and nonlinear scalar fields) the gravitational interaction in GR can have
the repulsion character, however, the PCS can not be solved by taking such systems into account in the
frame of GR \cite{mc4,mc5}. There were many attempts to resolve PCS in the frame of GR as well as other
classical theories of gravitation (see \cite{mc1,mc15} and Refs given herein). A number of regular
cosmological solutions was obtained in the frame of metric theories of gravitation and also other
theories, in the frame of which gravitation is described by using more general geometry than the
Riemannian one. In connection with this, note that the resolution of PCS means not only obtaining regular
cosmological solutions, but also excluding singular solutions of cosmological equations, as a result
generic feature of cosmological solutions has to be its regular character. Moreover, gravitation theory
and cosmological equations have to satisfy the correspondence principle with GR in the case of usual
gravitating systems with sufficiently small energy densities and weak gravitational fields excluding
nonphysical solutions. The greatest part of existent attempts to resolve the PCS does not satisfy
indicated conditions \cite{mc15}.

Usually the Universe evolution in the frame of standard cosmological scenario is considered beginning from
the time $t$ greater than the Planckian time $t_p\sim10^{-43} s$,when the energy density $\rho$ and
temperature $T$ were smaller than the Planckian ones: $\rho_p\sim 9\cdot 10^{112}\,
\frac{\text{J}}{\text{m}^3}$ and $T_p\sim 10^{32}\,K$.Then according to general opinion quantum
gravitational effects are negligible and gravitational field can be described classically. On the
contrary, at Planckian conditions ($\rho\ge\rho_p$, $T\ge T_p$) one means that quantum gravitational
effects must be essential, and possibly the PCS can be solved by taking these effects into account. So, in
order to solve the PCS the idea of quantum birth of the Universe was introduced
\cite{tryon,fomin}. There are several realizations of this idea in the frame of quantum cosmology by using
Wheeler-DeWitt equation (see review \cite{mc11}). The appearing closed micro-universe is transformed then into
macro-universe by virtue of  inflation. Although singular state with divergent energy density is absent in
such model, the problem of the beginning in the time of the Universe remains. The theory of quantum
micro-universe was developed further in the frame of loop quantum gravity. The space quantization permits
to continue solution for micro-universe of closed and flat type to the past and obtain bouncing solution
containing the compression stage before the expansion \cite{boj}. Note, this scenario has some vaguenesses,
in particular: if at compression stage the Universe was a macro-object filled by gravitating matter what
physical factors lead to transformation of macro-universe into micro-universe before a bounce, what is a
flat model with finite volume ?

The PCS was discussed also in the frame of other candidate for quantum gravity theory -– string theory/
M-theory \cite{lid,ven1,khory,mc12,mc13,ven2}. Some cosmological scenarios with evolution stage before the
Big Bang -– Pre-Big Bang scenario \cite{ven1} and ekpyrotic scenario \cite{khory} -– were proposed. Note
that HICM in a four-dimensional Einsteinian frame, generally speaking, are singular \cite{lid}. The obtaining
of  regular bouncing solutions is connected with violation of condition of energy density positivity for
gravitating matter . So, in Ref. \cite{ven1} non-local negative scalar field potential is introduced, in
Ref. \cite{ven2} a component  with negative energy density is used. To build oscillating cosmological
model, the specific negative scalar field potential was introduced in Refs. \cite{mc12,mc13}. Although the
energy density does not diverge by transition from compression to expansion, the singularity connected
with vanishing of the scale factor $R(t)$ remains.

As it was shown in a number of our papers (see \cite{mc15} and Refs given herein), the applying of gauge
approach to gravitation permits to build regular Big Bang theory by satisfying positivity property of
energy density for gravitating matter. The present paper is devoted to consideration of principal features
of developing approach and to further analysis of regular inflationary cosmology.  In Section 2
generalized cosmological Friedmann equations for HICM in the frame of gauge approach to gravitation are
introduced. In Section 3 these equations are applied to analyze inflationary cosmological models filled by
scalar fields and usual gravitating matter. As illustration of regular Big Bang theory in Section 4
particular inflationary cosmological model is discussed.

\section{II. GAUGE APPROACH TO GRAVITATION AND GENERALIZED COSMOLOGICAL FRIEDMANN EQUATIONS}

As it is known, the local gauge invariance principle is the basis of modern theory of fundamental physical
interactions. The theory of electro-week interaction, quantum chromodynamics, Grand Unified models of
particle physics were built by using this principle. From physical point of view, the local gauge
invariance principle establishes the correspondence between certain important conserving physical
quantities, connected according to the Noether's theorem with some symmetries groups, and fundamental
physical fields, which have as a source corresponding physical quantities and play the role of carriers of
fundamental physical interactions. The applying of this principle to gravitational interaction leads,
generally speaking, to generalization of Einsteinian theory of gravitation. Metric theories of gravitation
including GR, in the frame of which the energy-momentum tensor is a source of gravitational field, can be
introduced in the frame of gauge approach by using the localization of 4-parametric translation group
\cite{mc16} \footnote{Because in the frame of gauge approach the gravitational interaction is connected
with space-time transformations, the gauge treatment to gravitation has essential differences in
comparison with Yang-Mills fields connected with internal symmetries groups. As a result, there are
different gauge treatments of gravitational interaction not discussed in this paper.}. Because the
localized translation group is, in fact, the group of general coordinate transformations, from this point
of view the general covariance of GR plays dynamical role. At the same time the Lorentz group (group of
tetrad Lorentz transformations) in GR and other metric theories of gravitation does not play any dynamical
role from the point of view of gauge approach, because corresponding Noether's invariant in these theories
is identically equal to zero \cite{mc17}. If one means that the Lorentz group plays the dynamical role in
gauge field theory and the Lorentz gauge field exists in the nature, in this case we obtain with necessity
the gravitation theory in the Riemann-Cartan space-time (see, for example, \cite{hehl,hay}). Corresponding
theory is known as Poincare gauge theory of gravitation (PGTG). Gravitational field variables in PGTG are
the tetrad (translational gauge field) and Lorentz connection (Lorentz gauge field); corresponding field
strengths are torsion and curvature tensors. As sources of gravitational field in PGTG are covariant
generalizations of energy-momentum and spin tensors. Unlike gauge Yang-Mills fields, for which the
Lagrangian is quadratic in the gauge field strengths, gravitational Lagrangian of PGTG can include also
linear in curvature term (scalar curvature), which is necessary to satisfy the correspondence principle
with GR.

The first attempt to apply the simplest PGTG -– Einstein-Cartan theory -– in order to solve the PCS was
made in Refs. \cite{kop,traut}. By using some classical model for spinning matter, non-singular
cosmological solutions were obtained. However, it was shown later, these solutions have model character
and critically depend on spinning matter description; by another spin description (for example, by means
of Dirac field) the cosmological singularity does not disappear. Moreover, because in the case of spinless
matter the Einstein-Cartan theory is identical to GR, all singular solutions for Friedmannian HICM are
exact solutions of Einstein-Cartan theory of gravitation. The next step to apply the PGTG in order to
solve the PCS was made in Ref.\cite{mc14}. In the frame of PGTG based on general expression of
gravitational Lagrangian $L_G$ including both a scalar curvature and different invariants quadratic in the
curvature and torsion tensors gravitational equations for HICM were deduced, these equations lead to the
following generalized cosmological Friedmann equations (GCFE) \footnote {The structure of gravitational
equations of PGTG for HICM does not depend on detailed form of quadratic part of gravitational Lagrangian
$L_G$. As a result, these equations contain two independent parameters $f$ and $a$, which are linear
combinations of coefficients at terms of $L_G$ quadratic in the curvature and torsion tensors. The
mathematical requirement for cosmological equations for HICM to be differential equations of the same
order as in GR can be fulfilled if $a=0$ that leads to Eqs (1)-(2). Unlike metric theories of gravitation,
in PGTG terms of $L_G$ quadratic in the curvature tensor do not lead to high derivatives in GCFE.
Parameter $\alpha$ in Eqs. (\ref{1})--(\ref{2})is defined as
$\alpha=\frac{1}{3}\left(16\pi\right)^2f\,M_p^{-4}$. In our previous papers the parameter $\beta=-\alpha$
was used.}:

\begin{eqnarray}
\label{1}
 &\displaystyle{\frac{k}{R^2}+\left\{\frac{d}{dt}\ln\left[R\sqrt{\left|1+\alpha\left(\rho-
3p\right)\right|}\,\right]\right\}^2 \nonumber}\\
&\displaystyle{ =\frac{8\pi}{3M_p^2}\,\frac{\rho+
\frac{\alpha}{4}\left(\rho-3p\right)^2}{1+\alpha\left(\rho-3p\right)} \, ,}
\end{eqnarray}

\begin{eqnarray}
\label{2} &\displaystyle{\frac{\left[\dot{R}+R\left(\ln\sqrt{\left|1+\alpha\left(\rho-
3p\right)\right|}\,\right)^{\cdot}\right]^\cdot}{R}} \nonumber
 \\
&\displaystyle{=-\frac{4\pi}{3M_p^2}\,\frac{\rho+3p-\frac{\alpha}{2}\left(\rho-3p\right)^2}{
1+\alpha\left(\rho-3p\right)}\, ,}
\end{eqnarray}

where $k=+1,0,-1$ for closed, flat, open models respectively, $\alpha$ is indefinite parameter with
inverse dimension of energy density, $M_p$  is Planckian mass, a dot denotes differentiation with
respect to time. (The system of units with $\hbar=c=1$ is used). Note that Eqs.(\ref{1})--(\ref{2})
are valid also in the frame of the most general gauge theory of gravitation -- metric-affine theory
\cite{mc25,mc26}. From Eqs. (\ref{1})-(\ref{2}) follows the conservation law
 in usual form
\begin{equation}
\label{3} \dot{\rho}+3H\left(\rho+p\right)=0,
\end{equation}
where $H=\frac{\dot{R}}{R}$   is the Hubble parameter. Besides cosmological equations (\ref{1})--(\ref{2})
gravitational equations of PGTG lead to the following relation for the torsion function $S$
\begin{equation}
\label{4} S=-\frac{1}{4}\,\frac{d}{dt} \ln\left|1+\alpha(\rho-3p)\right|.
\end{equation}
Before we start discussing inflationary cosmology on a base of GCFE (\ref{1})-(\ref{2}), let us make
several remarks about these equations. At $\alpha\to 0$ the GCFE (\ref{1})-(\ref{2}) are transformed into
Friedmann cosmological equations of GR. The value of $\alpha^{-1}$ determines the scale of extremely high
energy densities. Solutions of GCFE (\ref{1})--(\ref{2}) coincide practically with corresponding solutions
of GR if the energy density is small $\left|\alpha(\rho-3p)\right|\ll 1$ ($p\neq\frac{1}{3}\rho$). The
difference between GR and PGTG can be essential at extremely high energy densities
$\left|\alpha(\rho-3p)\right|\gtrsim 1$. Ultrarelativistic matter ($p=\frac{1}{3}\rho$) and gravitating
vacuum ($p=-\rho$) with constant energy density are two exceptional systems because Eqs.
(\ref{1})--(\ref{2}) are identical to Friedmann cosmological equations of GR in these cases independently
on values of energy density and $S=0$. The behaviour of solutions of Eqs. (\ref{1})--(\ref{2}) depends
essentially on the following restriction on equation of state of gravitating matter at extreme conditions:
$p>\frac{1}{3}\rho$ or $p<\frac{1}{3}\rho$. Because of strong nucleon interaction we have for nuclear
matter $p>\frac{1}{3}\rho$ \cite{zel}; we will put that this restriction is valid also for gravitating
matter at extreme conditions, and the scale of extremely high energy densities determined by $\alpha^{-1}$
surpasses nuclear density. From physical point of view, we can suppose that the value of $\alpha^{-1}$ is
smaller than the Planckian energy density, but it will be shown later, the behaviour of solutions of GCFE
does not depend on this assumption.

The GCFE lead to restrictions on admissible values of energy density. In fact, if energy density $\rho$ is
positive and $\alpha>0$, from Eq.(\ref{1}) in the case $k=+1$, $0$ follows the relation:
\begin{equation}
\label{5} Z\equiv 1+\alpha\left(\rho-3p\right)\ge 0.
\end{equation}
The condition (\ref{5}) is valid not only for closed and flat models, but also for cosmological models of
open type ($k=-1$) (see below). In the case of models filled by gravitating matter with equation of state
$p=p(\rho)$, it is easy to obtain the solution of the system of Eqs. (\ref{1}) and (\ref{3}) in
quadratures \cite{mc25_}; these solutions are regular in metrics, Hubble parameter, its time derivative
and have bouncing character. The transition from compression to expansion takes place by reaching limiting
energy density defined by the following condition $Z=0$. At first the conclusion on possible existence of
limiting energy density for gravitating systems, close by which the gravitational interaction has the
character of repulsion, was obtained in Ref.\cite{mc14}. Later the idea on limiting energy density as
"universal law of the nature" was discussed by M.A. Markov \cite{mark}, and the value of limiting energy
density was postulated to be equal to the Planckian one. According to (\ref{5}) the value of limiting
energy density in our case depends on parameter $\alpha$ and equation of state for gravitating matter at
extreme conditions and can be essentially smaller than the Planckian one \footnote {The GCFE lead to
limiting energy density also in the case $\alpha<0$ if at extreme conditions $p<\frac{1}{3}\rho$
\cite{mc14}.}.

\section{III. Inflationary cosmological models and its properties }

Now by using GCFE (\ref{1})--(\ref{2}) we will study homogeneous isotropic models filled by interacting
scalar field $\phi$ minimally coupled with gravitation and gravitating matter with equation of state in
general form $p_m=p_m(\rho_m)$. (The generalization for the case with several scalar fields can be
made directly). Then the energy density $\rho$ and pressure $p$ take the form
\begin{equation}
\label{5_} \rho=\frac{1}{2}\dot{\phi}^2+V+\rho_m \quad (\rho>0), \quad p=\frac{1}{2}\dot{\phi}^2-V+p_m,
\end{equation}
where scalar field potential $V=V(\phi, \rho_m)$ includes the interaction between scalar field and
gravitating matter. In the most important particular case of radiation (ultrarelativistic matter)
the expressions of $V(\phi, \rho_m)$ can be obtained by taking into account temperature corrections
for given scalar field potentials \cite{mc27} and the following relation for energy density
$\rhm\sim T^4$. Because the form of equation of state for gravitating matter changes by
cosmological evolution and we do not know explicit form of scalar field potential, our analysis
will be made without its concretization. Our main aim will be to investigate properties of
inflationary cosmological solutions for early Universe. By using scalar field equation in
homogeneous isotropic space
\begin{equation}
\label{6} \ddot{\phi}+3H\dph=-\frac{\pat\,V}{\pat\,\phi}
\end{equation}
we obtain from Eqs. (\ref{3}), (\ref{5_}), (\ref{6}) the conservation law for gravitating matter
\begin{equation}
\label{7} \dot{\rho}_m\left(1+\frac{\pat\,V}{\pat\,\rhm}\right)+3H\left(\rhm+\prm\right)=0.
\end{equation}
By using Eqs.(\ref{5_})--(\ref{7}) the GCFE (\ref{1})--(\ref{2}) can be transformed to the following
form

\begin{widetext}
\begin{multline}
\label{8}
 \left\{
 H\left(
   Z+3\alpha
      \dph^2+\frac{3\alpha}{2}\,Y
   \right)
 +3\alpha\frac{\pat V}{\pat\phi}\dph\right\}^2
\\
+\frac{k}{R^2}\,Z^2 =\frac{8\pi}{3M_p^2}\,
 \left[
   \rhm+\frac{1}{2}\dph^2+V +\frac{1}{4}\alpha\,
   \left(4V-\dph^2+\rhm-3\prm\right)^2
 \right]
\,Z,
\end{multline}

{\allowdisplaybreaks
\begin{multline}
\label{9}
\dot{H}\left[
 Z+3\alpha
 \left(
   \dph^2+\frac{1}{2}\, Y
 \right)
\right]Z  +H^2 \left\{
 \left[
   Z-15\alpha\dph^2+\frac{3\alpha}{2}\, Y
   \right.\right.\\
   \left.\left.
   -\frac{9\alpha}{2}\,
        \frac{Y}{1+\frac{\pat V}{\pat\rhm}}
        \left(
            1+\frac{d\prm}{d\rhm}+\frac{\rhm+\prm}
                    {1+\frac{\pat V}{\pat\rhm}}\,
                \frac{\pat^2 V}{\pat\rhm^2}
        \right)
         -\frac{9\alpha}{2}\,
            \frac{\left(\rhm+\prm\right)^2}{\left(1+\frac{\pat V}{\pat\rhm}\right)^2}\,
            \frac{d^2}{d\rhm^2}
            \left(3\prm-4V\right)
 \right]Z
    \right.\\
    \left.
\vphantom{\frac{\tst \rhm+\prm}{\tst 1+\frac{\pat V}{\pat \rhm}}}
     -18\alpha^2\,
 \left(\dph^2+\frac{1}{2}Y \right)^2
\right\}
-12\alpha H\dph \left\{
    3\alpha\frac{\pat V}{\pat\phi}
    \left(\dph^2+\frac{1}{2}Y\right)
    \right.\\
    \left.
    +
    \left[
        \frac{\pat V}{\pat\phi}+\frac{9}{8}\,
        \frac{\tst \rhm+\prm}{\left(\tst 1+\frac{\pat V}{\pat \rhm}\right)^2}\,
        \frac{\pat^2 V}{\pat\phi\pat\rhm}
        \left(1+\frac{1}{3}\ddrhm\left(\prm+2V\right)\right)
    \right]Z
\right\}
\\
+3\alpha \left[
    \frac{\pat^2 V}{\pat\phi^2}\dph^2-\left(\frac{\pat V}{\pat\phi}\right)^2
\right]Z -18\alpha^2\left(\frac{\pat V}{\pat\phi}\right)^2\dph^2
\\
=\frac{8\pi}{3M_p^2} \left[
    V-\dph^2-\frac{1}{2}\left(\rhm+3\prm\right)
    +\frac{1}{4}\alpha\left(4V-\dph^2+\rhm-3\prm\right)^2
\right]Z,
\end{multline}
 }
\end{widetext}

where $Z=1+\alpha\left(4V-\dph^2+\rhm-3\prm\right)$ and $Y=\frac{\tst \rhm+\prm}{\tst 1+\frac{\pat V}{\pat \rhm}}\,
\left( \ddrhm\,\left( 3\prm-4V \right)-1 \right)$. The formula (\ref{4}) takes the form
\begin{equation}
\label{10} S=-\frac{3\alpha}{2Z}\, \left[
    H\left(\dph^2+\frac{1}{2}Y\right)
    +\frac{\pat V}{\pat \phi}\dph
\right].
\end{equation}
The relation (\ref{5}) determines the following restriction on admissible values of variables for scalar
field and gravitating matter:
\begin{equation}
\label{11} \qquad \dph^2\le4V+\alpha^{-1}+\rhm-3\prm.
\end{equation}
Now let us introduce the 3-dimensional space $P$ with axes $(\phi,\dot{\phi},\rho_m)$. The domain of
admissible values of scalar field $\phi$, time derivative $\dot{\phi}$ and energy density $\rho_m$ in
space $P$ determined by (\ref{11}) is limited by bound $L$ defined as
\begin{equation} \label{12}
Z=0\quad \text{or}\quad \dot\phi=\pm\left(4V+\alpha^{-1}+\rhm-3\prm\right)^{\frac{1}{2}}.
\end{equation}
From Eq. (\ref{8}) the Hubble parameter on the bound $L$  is equal to
\begin{equation}
\label{13} H_{L}=-\frac{\dst \frac{\pat V}{\pat \phi}\, \dot\phi}{ \dst \dph^2+\frac{1}{2}Y}.
\end{equation}

Let us consider the most important general properties of cosmological solutions of GCFE
(\ref{8})--(\ref{9}). At first note, by given initial conditions for variables ($\phi$,
$\dot{\phi},\rho_m$) and also in the case $k=\pm 1$ for $R$ there are two different solutions
corresponding to two values of the Hubble parameter following from Eq.~(\ref{8}):
\begin{equation}
\label{14} H_{\pm}=\frac{\dst -3\alpha\,\frac{\pat V}{\pat \phi}\, \dot\phi \pm \sqrt{D}}{ \dst
Z+3\alpha[\dph^2+\frac{1}{2}Y]},
\end{equation}
where
\begin{multline}
\label{15} D=\frac{8\pi}{3M_p^2}\,[\rhm+\frac{1}{2}\dph^2+V \\
 +\frac{1}{4}\alpha\,
\left(4V-\dph^2+\rhm-3\prm\right)^2]\,Z-\frac{k}{R^2}Z^2.
\end{multline}
Obviously, the expression (\ref{14}) for $H_{\pm}$ will be regular, if $Y\ge 0 $. This relation is
valid, in particular, for all models filled by gravitating matter with $p_m \ge \frac{\rho_m}{3} $
and scalar fields with potentials applying in chaotic inflation theory. Unlike GR, the values of
$H_{+}$ and $H_{-}$ in GTG are sign-variable and, hence, both solutions corresponding to $H_{+}$
and $H_{-}$ can describe the expansion as well as the compression in dependence on its sign. Below
we will call solutions of GCFE corresponding to $H_{+}$ and $H_{-}$ as $H_{+}$-solutions and
$H_{-}$-solutions respectively. Note that at asymptotics like GR the sign of $H_{-}$ is negative
and the sign of $H_{+}$ is positive.  In points of bound $L$ we have $D=0$, $H_{+}=H_{-}$ and the
Hubble parameter is determined by~(\ref{13}). If initial conditions correspond to asymptotics of
$H_{-}$-solution, then unlike GR by compression stage the derivative $\dot{\phi}$ does not diverge
and by reaching the bound $L$ the transition from $H_{-}$-solution to $H_{+}$-solution takes place.
In fact, by using the following formula for $H_{\pm}$-solutions
 \begin{equation}
 \label{16}
\dot{Z}=6\alpha\,
    \frac{\dst \frac{\pat V}{\pat \phi}\, \dot\phi\, Z \pm
                \sqrt{D}\left(\dph^2+\frac{1}{2}Y \right)}
        { \dst Z+3\alpha\left(\dph^2+\frac{1}{2}Y\right)}
\end{equation}
it is easy to show that
\begin{equation}\label{17}
\lim_{Z\to 0} \dot{H}_{\pm}=
    \dot{H}_L+\frac{2\pi}{3\alpha M_p^2}\,
        \frac{\tst \rhm+ \prm +\dph^2 }{\tst \dph^2+\frac{1}{2}Y}.
 \end{equation}
From (\ref{17}) follows that in points of bound $L$ the derivatives $\dot{H}_{+}$ and $\dot{H}_{-}$
are equal and its values do not depend on the model type, as a result we have the smooth transition
from $H_{-}$-solution to $H_{+}$-solution on bound $L$, and corresponding cosmological solutions
for all types models are regular in metrics, Hubble parameter and its time derivative.  Note, that
in points of bound $L$ conditions of uniqueness of solutions of Eqs. (\ref{8})-(\ref{9}) are not
fulfilled, as a result there are specific solutions \cite{mc3,mc15}, trajectories of which are
situated on the bound $L$ and have with $H_{\pm}$-solutions common points, $H_{-}$--solutions reach
the bound $L$ and $H_{+}$-solutions originate from them, the surface $Z=0$ is envelope in space $P$
for cosmological solutions \footnote{From mathematical point of view the appearance of specific
solutions is connected with the fact, that the Eqs. (1)--(2) were multiplied by $Z^2$ when its
transformation to the (9)--(10) was performed. Specific solutions can be excluded by corresponding
transformation of Eq. (10).}. According to Eq. (\ref{16}) and (\ref{4}) the function $S$ has the
following asymptotics for $H_{+}$- and $H_{-}$-solutions at $Z\to 0$:
\begin{equation}\label{18}
\lim_{Z\to 0}S=-\frac{1}{4}\,\lim_{Z\to 0}\frac{\dot{Z}}{Z}\sim\mp Z^{-\frac{1}{2}}.
\end{equation}
Unlike flat and open models, for which $H_{+}= H_{-}$ only in points of bound $L$ and regular inflationary
models include $H_{+}$-- and $H_{-}$--solutions reaching bound $L$, in the case of closed models the
regular transition from $H_{-}$-solution to $H_{+}$-solution is possible without reaching the bound $L$.
It is because by certain value of $R$ according to (\ref{14})--(\ref{15}) we have $H_{+}= H_{-}$ in the
case $Z\neq 0$. Such models are regular also in torsion . Regular inflationary solution
of such type was considered in Ref.\cite{mc6}.

All discussed cosmological solutions have bouncing character, but in the presence of scalar fields a
bounce takes place not in points of bound $L$ ($Z=0$). In order to study the behaviour of cosmological
models at a bounce, let us analyze extreme points for the scale factor $R(t)$: $R_0=R(0)$, ${H_0=H(0)=0}$.
(This means that in the case of $H_{+}$--solutions $H_{+0}=0$ and in the case of $H_{-}$--solutions
$H_{-0}=0$). Denoting values of quantities at $t=0$ by means of index "0", we obtain from
(\ref{8})--(\ref{9}):

\begin{widetext}
\begin{equation}
\label{19} \frac{k}{R_0^2}\,Z_0^2 +9\alpha^2{\left(\frac{\pat V}{\pat\phi}\right)\!}_0^2\dph_0^2
=\frac{8\pi}{3M_p^2}\,
 \left[
   \rhmz+\frac{1}{2}\dph_0^2+V_0 +\frac{1}{4}\alpha\,
   \left(4V_0-\dph_0^2+\rhmz-3\prmz\right)^2
 \right]
\,Z_0,
\end{equation}

\begin{multline}
\label{20} \dot{H}_0= \left\{
    \frac{8\pi}{3M_p^2}
    \left[
        V_0-\dph_0^2-\frac{1}{2}\left(\rhmz+3\prmz\right)
        +\frac{1}{4}\alpha\left(4V_0-\dph_0^2+\rhmz-3\prmz\right)^2
    \right]
    \right.\\
    \left.
    -3\alpha
    \left[
        {\left(\frac{\pat^2 V}{\pat\phi^2}\right)\!}_0\dph_0^2
        -{\left(\frac{\pat V}{\pat\phi}\right)\!}_0^2
    \right]
    +18\alpha^2{\left(\frac{\pat V}{\pat\phi}\right)\!}_0^2\dph_0^2\:Z_0^{-1}
\right\}
 \left\{
 Z_0+3\alpha
 \left[
   \dph_0^2+\frac{1}{2}Y_0
 \right]
\right\}^{-1},
\end{multline}

\noindent where $Z_0=1+\alpha\left(4V_0-\dot{\phi}_0^2+\rho_{m0}-3p_{m0}\right)$. A bounce point is
described by Eq. (\ref{19}), if the value of $\dot{H}_0$ is positive. By using Eq.(\ref{19}) we can
rewrite the expression of $\dot{H}_0$ in the form

\begin{eqnarray}
&\displaystyle{  \dot{H}_0= \left\{
    \frac{8\pi}{M_p^2}
    \left[
        V_0+\frac{1}{2}\left(\rhmz-\prmz\right)
        +\frac{1}{4}\alpha\left(4V_0-\dph_0^2+\rhmz-3\prmz\right)^2
    \right]\nonumber
    \right.}\\
    &\displaystyle{
    \left.
    -3\alpha
    \left[
        {\left(\frac{\pat^2 V}{\pat\phi^2}\right)\!}_0\dph_0^2
        -{\left(\frac{\pat V}{\pat\phi}\right)\!}_0^2
    \right]
    -\frac{2k}{R_0^2}\,Z_0
\right\}
  \left\{
 Z_0+3\alpha
 \left[
   \dph_0^2+\frac{1}{2}Y_0
 \right]
\right\}^{-1}\!.}\label{21}
\end{eqnarray}
\end{widetext}

We see from (\ref{21}) unlike GR the presence of gravitating matter satisfying the energy dominance
condition ($p_m\le \rho_m$) does not prevent from the bounce realization\footnote{In GR a bounce is
possible only in closed models if the following condition
$V_0-\dot{\phi}_0^2-\frac{1}{2}(\rho_{m0}+3p_{mo})>0$ takes place.}. Eq. (\ref{19}) determines in
space $P$ extremum surfaces depending on the value of $\alpha$ and in the case of closed and open
models also parametrically on the scale factor~$R_0$. In the case of various scalar field
potentials applying in inflationary cosmology the value of $\dot{H}_{+0}$ or $\dot{H}_{-0}$ is
positive on the greatest part of extremum surfaces, which can be called "bounce
surfaces"\footnote{If $\alpha^{-1}\ll M_p^4$ the derivative $\dot{H}_0$ is negative in the
neighbourhood of origin of coordinates in space $P$ that leads to appearance of oscillating
solutions of GCFE\cite{mc28}.}. By giving concrete form of potential $V$ and choosing values
of~$R_0$, $\phi_0$, $\dot{\phi}_0$ and $\rho_{m0}$ at a bounce, we can obtain numerically
particular bouncing solutions of GCFE for various values of parameter $\alpha$. Like GR, if initial
value of scalar field at the beginning of cosmological expansion is not small ($\phi \ge 1 M_p$)
\cite{mc27}, corresponding solution is inflationary cosmological solution containing in addition to
inflationary stage also compression stage, transition stage from compression to expansion and a
stage after inflation with oscillating regime for scalar field. For given scalar field potential
properties of regular inflationary cosmological solutions depend on initial conditions at a bounce
and parameter $\alpha$. Numerical analysis of such solutions in the case of simplest scalar field
potentials applying in chaotic inflation was carried out in Ref. \cite{mc29}. Note that by equal
initial conditions characteristics of inflationary stage in developing thery coincide with that of
GR \cite{mc29}.

The analysis of GCFE shows, that some properties of cosmological solutions depend essentially on
parameter $\alpha$, i.e. on the scale of extremely high energy densities. From physical point of view
interesting results can be obtained, if the value of $\alpha^{-1}$   is much less than the Planckian
energy density\cite{mc6}, i.e. in the case of large values of parameter $\alpha$ (by imposing $M_p=1$). In
order to investigate cosmological solutions at the beginning of cosmological expansion in this case, let
us consider the GCFE by supposing that
\begin{eqnarray}
\label{22} &\displaystyle{\left|\alpha\left(4V-\dph^2+\rhm-3\prm\right)\right|\gg 1,\nonumber}
\\
&\displaystyle{\rhm+\frac{1}{2}\dph^2+V\ll \alpha \left(4V-\dph^2+\rhm-3\prm\right)^2.}
\end{eqnarray}
Note that the second condition (\ref{22}) does not exclude that ultrarelativistic matter energy density
can dominate at a bounce. We obtain:

\begin{widetext}
{\allowdisplaybreaks
\begin{multline}
\label{23}
\frac{k}{R^2}\left(4V-\dph^2+\rhm-3\prm\right)^2+
\left\{H
    \left[
        4V+2\dph^2+\rhm-3\prm
        +\frac{3}{2}Y
    \right]
    \right.\\
    \left.
    \vphantom{\frac{\tst \rhm+\prm}{\tst 1+\frac{\pat V}{\pat \rhm}}}
    +3\frac{\pat V}{\pat\phi}\dph
\right\}^2 =\frac{2\pi}{3M_p^2}\,\left(4V-\dph^2+\rhm-3\prm\right)^3,
\end{multline}

\begin{multline}
\label{24} 
\dot{H} \left(
    4V+2\dph^2+\rhm-3\prm
    +\frac{3}{2}Y
\right)
\left(4V-\dph^2+\rhm-3\prm\right) +H^2 \left\{
    \left[
        4V-16\dph^2+\rhm-3\prm
        \vphantom{\frac{\tst \rhm+\prm}{\tst 1+\frac{\pat V}{\pat \rhm}}}
        \right.\right.\\
        \left.\left.
        +\frac{3}{2}Y
        -\frac{9}{2}\,
        \frac{Y}{1+\frac{\pat V}{\pat\rhm}}
        \left(
            1+\frac{d\prm}{d\rhm}+\frac{\rhm+\prm}
                    {1+\frac{\pat V}{\pat\rhm}}\,
                \frac{\pat^2 V}{\pat\rhm^2}
        \right)
         -\frac{9}{2}\,
            \frac{\left(\rhm+\prm\right)^2}{\left(1+\frac{\pat V}{\pat\rhm}\right)^2}\,
            \frac{d^2}{d\rhm^2}
            \left(3\prm-4V\right)
    \right]
    \right.\\
    \left.
  \times
 \left(4V-\dph^2+\rhm-3\prm\right)
    -18\left(\dph^2+\frac{1}{2}Y\right)^2
\right\}
\\
-12H\dph \left\{
    \left[
        \frac{\pat V}{\pat\phi}+\frac{9}{8}\,
        \frac{\tst \rhm+\prm}{\left(\tst 1+\frac{\pat V}{\pat \rhm}\right)^2}\,
        \frac{\pat^2 V}{\pat\phi\pat\rhm}
        \left(1+\frac{1}{3}\ddrhm\left(\prm+2V\right)\right)
    \right]
    \right.\\
    \left.\times
    \left(4V-\dph^2+\rhm-3\prm\right)
    +3\frac{\pat V}{\pat\phi}
    \left(\dph^2+\frac{1}{2}Y\right)
\right\}
\\
+3 \left[
    \frac{\pat^2 V}{\pat\phi^2}\dph^2-\left(\frac{\pat V}{\pat\phi}\right)^2
\right]\left(4V-\dph^2+\rhm-3\prm\right)
\\
-18\left(\frac{\pat V}{\pat\phi}\right)^2\dph^2 =\frac{2\pi}{3M_p^2} \left(
    4V-\dph^2+\rhm-3\prm
\right)^3.
\end{multline}
}

According to Eq.~(\ref{23}) the Hubble parameter in considered approximation is equal to

\begin{multline}
\label{25}
H_{\pm}= \left[
    4V+2\dph^2+\rhm-3\prm +\frac{3}{2}Y
\right]^{-1}
\\
\!\times \left[
    -3\frac{\pat V}{\pat \phi}\, \dph\pm
    \left|4V-\dph^2+\rhm-3\prm\right|
    \sqrt{
        \frac{2\pi}{3M_p^2}\left(4V-\dph^2+\rhm-3\prm\right)
        -\frac{k}{R^2}
    }\,\,
\right]
\end{multline}
 \end{widetext}

and  extreme points of the scale factor are determined by the following condition
\begin{eqnarray}
\label{26} \frac{k}{R_0^2}+9 \left[
    \frac{{\left(\dst \frac{\pat V}{\pat\phi}\right)\!}_0\dph_0}
        {\dst 4V_0-\dph_0^2+\rhmz-3\prmz}
\right]^2 \nonumber \\
 =\frac{2\pi}{3M_p^2}\,\left(4V_0-\dph_0^2+\rhmz-3\prmz\right).
\end{eqnarray}
From Eq. (\ref{24}) the time derivative of the Hubble parameter at extreme points is
 {\allowdisplaybreaks
\begin{multline}
\label{27} \dot{H}_0 = \left\{
    \frac{2\pi}{3M_p^2} \left(4V_0-\dph_0^2+\rhmz-3\prmz\right)^2
    \right.\\
    \left.
    -3
    \left[
       {\left(\frac{\pat^2 V}{\pat\phi^2}\right)\!}_0\dph_0^2
       -{\left(
            \frac{\pat V}{\pat\phi}
        \right)\!}_0^2%
    \right]
    +\frac{18{\left(\frac{\pat V}{\pat\phi}\right)\!}_0^2\dph^2}{
        4V_0-\dph_0^2+\rhmz-3\prmz}
\right\}
\\
\times \left[
    4V_0+2\dph_0^2+\rhmz-3\prmz+\frac{3}{2}Y_0
\right]^{-1}.
\end{multline}
 }

From Eqs.(\ref{23})--(\ref{27}) follows that in approximation (\ref{22}) the dynamics of inflationary
cosmological models at a bounce does not depend on parameter $\alpha$. In the case of models containing at
a bounce ultrarelativistic matter and scalar fields, the dynamics in considered approximation does not
depend on ultrarelativistic matter, if interaction terms in scalar field potential can be neglected.

The analysis given in this Section shows that existence of limiting bound $L$ and bounce surface in
space $P$ ensures regular character of inflationary cosmological solutions in metrics, Hubble
parameter and its time derivative. Characteristic feature of inflationary cosmological models is
the presence of the stage of regular transition from compression to expansion. The duration of this
stage is several order smaller than duration of inflationary stage. If we take into account that
duration of inflationary stage in chaotic inflation is extremely small \cite{mc27}, we can tell
that our inflationary models correspond to regular Big Bang or Big Bounce.

\section{IV. Particular regular inflationary cosmological model}

\begin{figure*}[htb!]
\begin{minipage}{0.48\textwidth}
\includegraphics{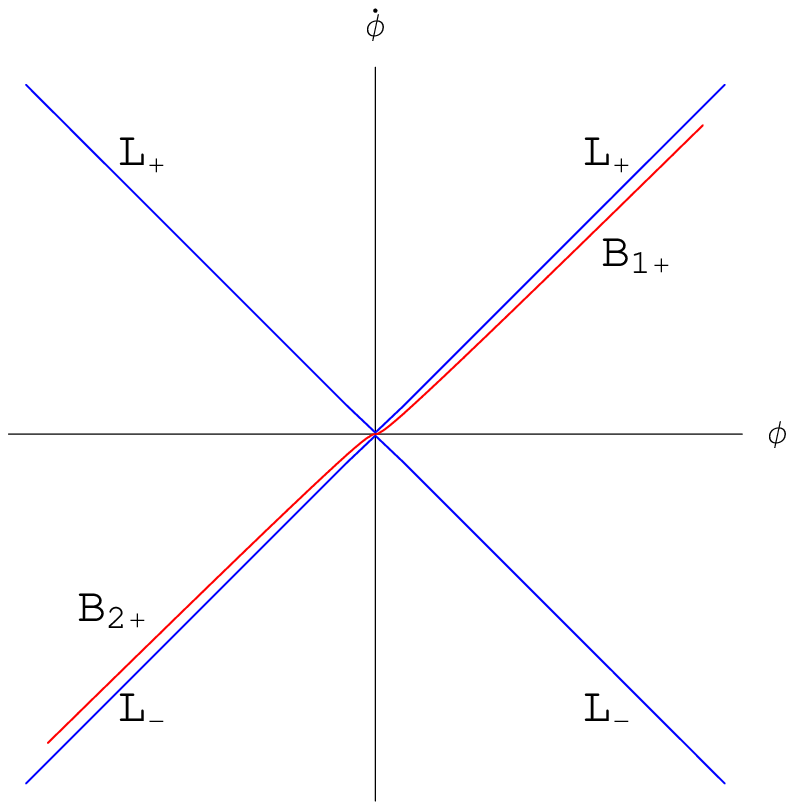}%
\end{minipage}\, \hfill\,
\begin{minipage}{0.48\textwidth}
\includegraphics{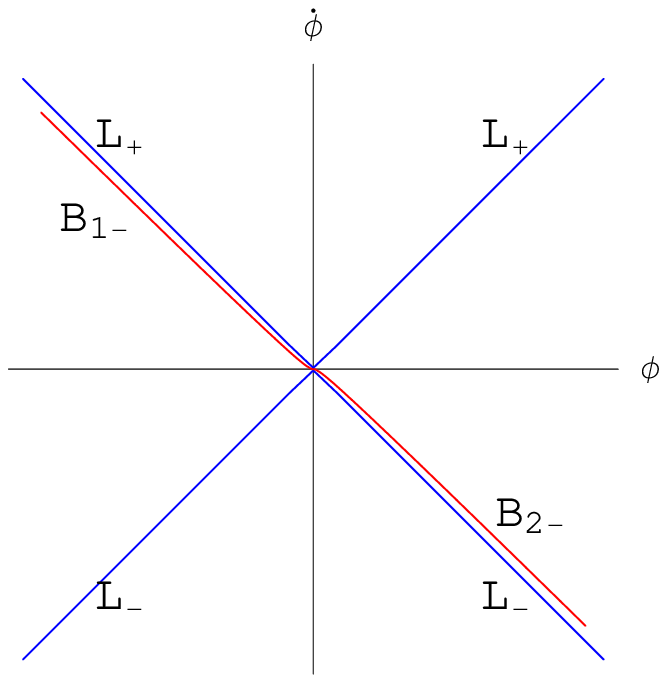}%
\end{minipage}
\caption[]{\label{figm1} Bound $L_\pm$-curves and bounce curves for flat models in the case of potential
$V=\frac{1}{2}m^2\phi^2$.}
\end{figure*}

As illustration of discussed theory regular cosmological models in the simplest particular case
will be considered in this Section. We will consider models including noninteracting scalar field
with potential $V(\phi)$ and ultrarelativistic matter ($\prm=\frac{1}{3}\rhm$)\cite{mc2}. In this
case the bound $L$ in space $P$ is reduced to two cylindric surfaces
$\dot{\phi}=\pm\left(4V+\alpha^{-1}\right)^{\frac{1}{2}}$. Bounce surface is reduced also to
cylindric surfaces in the case under consideration, when the scale of extremely high energy
densities is much smaller than the Planckian energy density (see (\ref{26})). In connection with
this we will consider instead of space $P$ the plane of variables ($\phi$, $\dot\phi$) and
intersections of the bound $L$ and bounce surface with this plane. We have in this plane two bound
$L_\pm$--curves and in the case of flat models two bounce curves $B_1$ and $B_2$
determined~by~equation \footnote{The neighbourhood of origin of coordinates is not considered in
this approximation, the behavior of extremum curves near origin of coordinates was examined in
Ref.~\cite{mc28,mc29}.}
\[
4V_0-\dot{\phi}_0^2=3\left(\frac{M_p^2}{2\pi}\,{V_0'}^{2}\,\dot{\phi}_0^2\right)^{\frac{1}{3}}\!.
\]
Each of two curves $B_{1,2}$ contains two parts corresponding to vanishing of $H_{+}$ or $H_{-}$ and
denoting by ($B_{1+}$, $B_{2+}$) and ($B_{1-}$, $B_{2-}$) respectively. If $V'$ is positive (negative) in
quadrants 1 and 4 (2 and 3) on the plane ($\phi$, $\dot\phi$), the bounce will take place in points of
bounce curves $B_{1+}$ and $B_{2+}$ ($B_{1-}$ and $B_{2-}$) in quadrants 1 and 3 (2 and 4) for
$H_{+}$-solutions ($H_{-}$-solutions) (see Fig.~\ref{figm1}).
To analyze flat bouncing models we have to take into account  that besides regions lying between curves
$L_\pm$ and corresponding bounce curves the sign of values $H_{+}$ and $H_{-}$ for applying potentials is
normal: $H_{+}>0$, $H_{-}<0$. The Hubble parameter $H_{+}$ is negative in regions between curves ($L_{+}$
and $B_{1+}$), ($L_{-}$ and $B_{2+}$), and the value of $H_{-}$ is positive in regions between curves
($L_{+}$ and $B_{1-}$), ($L_{-}$ and $B_{2-}$). As it was noted above any cosmological solution has to
contain both $H_{-}$- and $H_{+}$-solution and regular transition from $H_{-}$-solution to
$H_{+}$-solution takes place in points of $L_{\pm}$ where $H_{+}=H_{-}$. In the case of open and closed
models Eq.(\ref{26}) determines 1-parametric family of bounce curves with parameter $R_0$. Bounce curves
of closed models are situated
 in region between two bounce curves $B_{1}$ and $B_{2}$ of
flat models, and in the case of open models bounce curves are situated in two regions between the curves:
$L_{+}$ and $B_{1}$, $L_{-}$ and $B_{2}$. In general case, when approximation (\ref{22}) is not valid,
bounce surface in space $P$ of cosmological models including scalar field and ultrarelativistic matter
determined by Eq.(\ref{19}) in space $P$ depends on parameter $\alpha$ and it is not more cylindric
surface. The situation concerning cosmological solutions of Eqs. (\ref{8})--(\ref{9}) does not change.

\begin{figure*}[htb!]
\begin{minipage}{0.48\textwidth}
\includegraphics{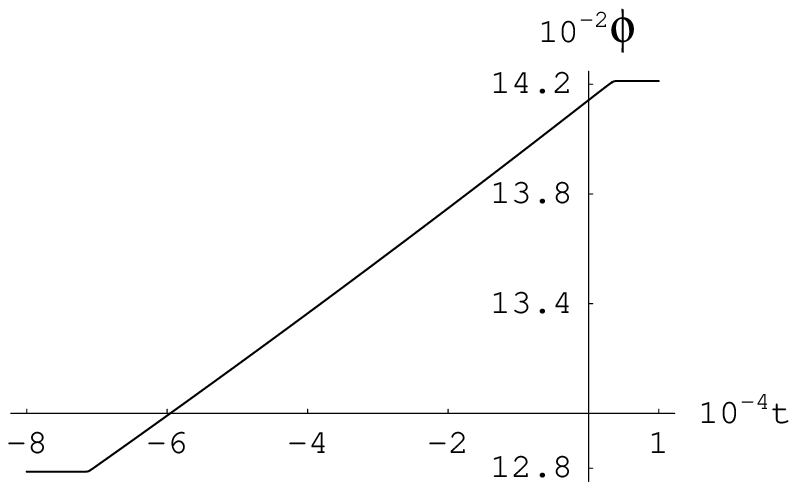}%
\end{minipage}\, \hfill\,
\begin{minipage}{0.48\textwidth}
\includegraphics{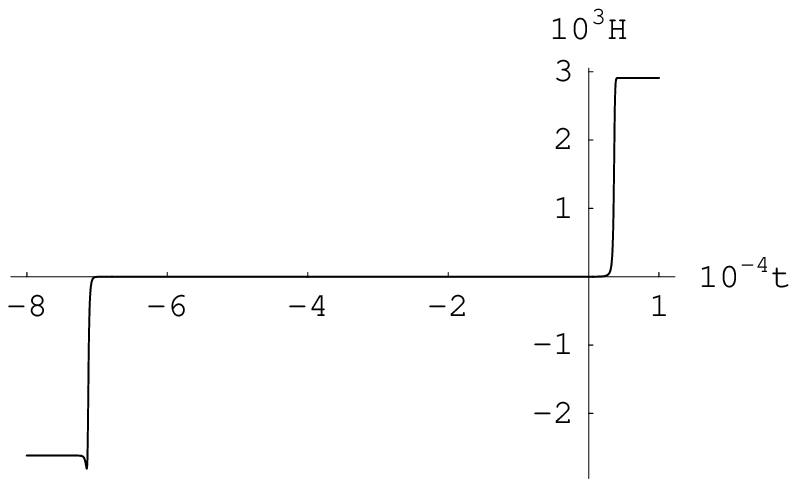}%
\end{minipage}
\caption{\label{figm2}The stage of transition from compression to expansion.}
\end{figure*}

\begin{figure*}[htb!]
\begin{minipage}{0.48\textwidth}
\includegraphics{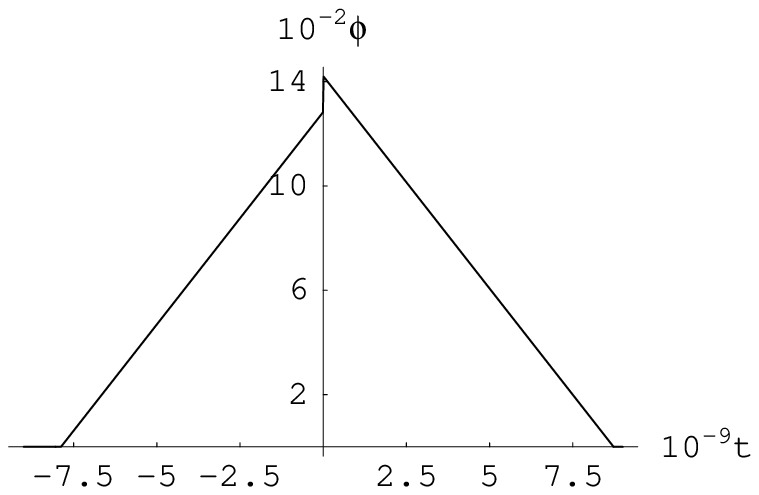}%
\end{minipage}\, \hfill\,
\begin{minipage}{0.48\textwidth}
\includegraphics{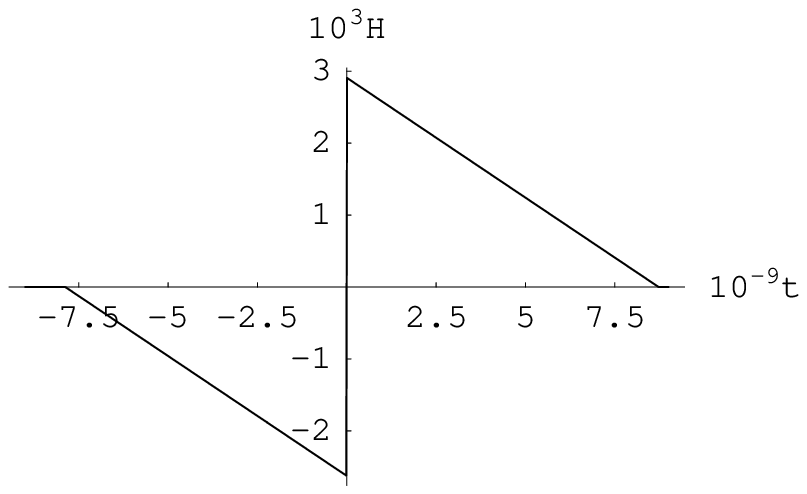}%
\end{minipage}
\caption{\label{figm3}Quasi-de-Sitter stage of compression and inflationary stage.}
\end{figure*}

\begin{figure*}[htb!]
\begin{minipage}{0.48\textwidth}
\includegraphics{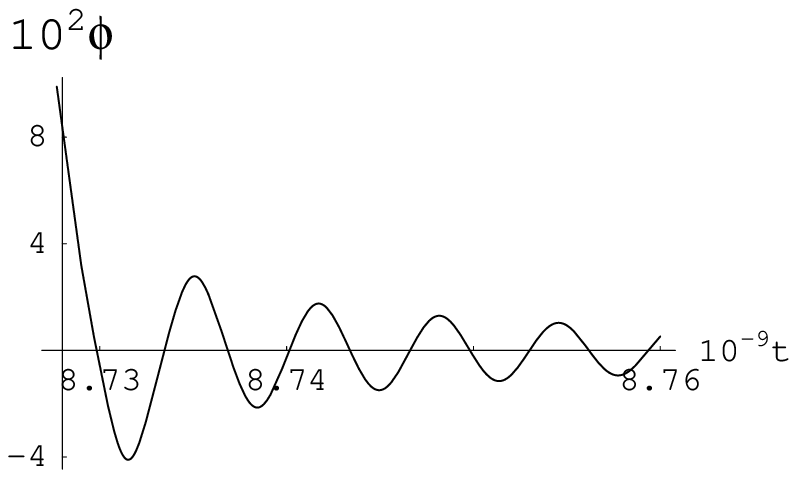}%
\end{minipage}\, \hfill\,
\begin{minipage}{0.48\textwidth}
\includegraphics{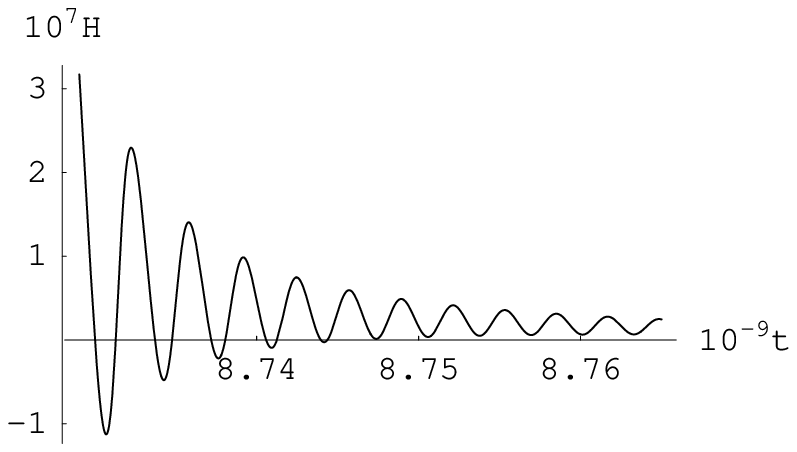}%
\end{minipage}
\caption{\label{figm4}The stage after inflation.}
\end{figure*}

 Below particular bouncing cosmological inflationary solution for flat model by using scalar field
potential in the form $V=\frac{1}{2}m^2\phi^2$ (${m=10^{-6}M_p}$)is given. The solution was obtained by
numerical integration of Eqs. (\ref{6}), (\ref{9}) and by choosing in accordance with Eq.(\ref{19}) the
following initial conditions at a bounce: $\phi_0=\sqrt{2}\, 10^3\, M_p$, $\dot{\phi}_0=\sqrt{3.96757
V_0\,}$ ($\alpha=10^{14}M_p^{-4}$); radiation energy density is negligibly small, initial value of $R_0$
can be arbitrary.  The dynamics of the Hubble parameter and scalar field is presented for different stages
of obtained bouncing solution in Figs. \ref{figm2}--\ref{figm4} (by choosing $M_p=1$).
The transition stage from compression to expansion (Fig.~\ref{figm2}) is essentially asymmetric with
respect to the point $t=0$ because of $\dot{\phi}_0\neq 0$. In course of transition stage the Hubble
parameter changes from maximum in module negative value at the end of compression stage to maximum
positive value at the beginning of expansion stage. The scalar field changes linearly, the derivative
$\dot\phi$ grows at first being positive to maximum value $\dot{\phi}\sim\dot{\phi}_0$ and then the value
of $\dot\phi$ decreases and becomes negative. In the case of presence of radiation, which can give the
main contribution to energy density at a bounce, the duration of transition stage changes. Quasi-de-Sitter
inflationary stage and quasi-de-Sitter compression stage are presented in Fig.~\ref{figm3}. As was noted
above, characteristics of inflationary stage do not depend practically on parameter $\alpha$ and coincide
with that of GR. The amplitude and frequency of oscillating scalar field after inflation
(Fig.~\ref{figm4}) are close to that of GR, however, the behaviour of the Hubble parameter after inflation
in considering case with large value of parameter $\alpha$ is essentially noneinsteinian, at first the
Hubble parameter oscillates near the value $H=0$, and later the Hubble parameter becomes positive and
decreases with the time like in GR. Before quasi-de-Sitter compression stage there are also oscillations
of the Hubble parameter and scalar field not presented in Figs. \ref{figm2}--\ref{figm4}.
Ultrarelativistic matter, which could dominate at a bounce has negligibly small energy densities at
quasi-de Sitter stages. At the same time the gravitating matter could be at compression stage in more
realistic bouncing models. As it follows from our consideration regular character of such inflationary
cosmological models has to be ensured by cosmological equations of PGTG.

The interaction between scalar fields and radiation leads to quantitative corrections of considered
cosmological models. In accordance with Eq. (\ref{7}) temperature corrections for scalar field potentials
change the connection between scale factor $R$ and radiation energy density and can be also essential for
more late stages of cosmological evolution, when energy densities are sufficiently small and consequences
of PGTG and GR coincide. In particular, these corrections can give additional contribution to the effect
of acceleration of cosmological expansion (if relic scalar fields exist).

Note that bouncing character have solutions not only in classical region, where scalar field potential,
kinetic energy density of scalar field and energy density of gravitating matter do not exceed the
Planckian energy density, but also in regions, where classical restrictions are not fulfilled and
according to accepted opinion quantum gravitational effects can be essential.

\section{V. Conclusion}

As we see, the applying of gauge approach to gravitational interaction permits to build consequent
field theoretical scheme in the frame of 4-dimensional physical space-time, which is free of
principal difficulties of GR by description of early Universe. Satisfying the correspondence
principle with GR in the case of gravitating systems with rather small energy densities,
generalized cosmological Friedmann equations lead to conclusion, that at extreme conditions
gravitational interaction has the repulsion character in the case of usual gravitating systems with
positive values of energy density satisfying the energy dominance condition. This means, there is
not necessity to refuse fundamental physical requirement of energy density positivity for physical
matter. The solution of PCS is obtained by classical description of gravitational field without
quantum gravitational corrections. Moreover, from the point of view of developing approach, the
Planckian era could be absent by evolution of our Universe. Unlike loop quantum cosmology, the
Universe is macro-object at all stages of its evolution including the transition stage from
compression to expansion. To build a realistic cosmological model we have to know the content and
properties of gravitating matter at different stages of its evolution. It is of principal interest
to investigate physical processes at the beginning of cosmological expansion depending on limiting
energy density and limiting temperature, and to obtain observable physical consequences depending
on parameter $\alpha$. From physical point of view, it is interesting also the building of
cosmological models by breakdown of its homogeneity and isotropy.



%





\end{document}